\documentclass{mem}
\usepackage{natbib}\usepackage{txfonts}\usepackage{balance}
\usepackage{graphicx}
\usepackage[a4paper,breaklinks,dvipdfm]{hyperref}
\idline{000}{001}
\begin{document}
\def\teff{$T\rm_{eff }$}
\def\kms{$\mathrm {km s}^{-1}$}

\title{
Fundamental Cosmology in the E-ELT Era
}

   \subtitle{}

\author{
C.\ J.\ A.\ P.\ Martins\inst{1}, M.\ C.\ Ferreira\inst{1,2}, M.\ D.\ Juli\~ao\inst{1,3}, A.\ C.\ O.\ Leite\inst{1,2}, A.\ M.\ R.\ V.\ L.\ Monteiro\inst{1,2,4}, P.\ O.\ J.\ Pedrosa\inst{1,2} and P.\ E.\ Vielzeuf\inst{1,2}
          }

  \offprints{C. Martins}

\institute{
Centro de Astrof\'{i}sica da Universidade do Porto, Rua das Estrelas, 4150-762 Porto, Portugal
\email{Carlos.Martins@astro.up.pt}
\and
Faculdade de Ci\^encias, Universidade do Porto, Rua do Campo Alegre 687, 4169-007 Porto, Portugal
\and
Faculdade de Engenharia, Universidade do Porto, Rua Dr Roberto Frias, 4200-465 Porto, Portugal
\and
Department of Applied Physics, Delft University of Technology, P.O. Box 5046, 2600 GA  Delft, The Netherlands
}

\authorrunning{Martins {\it et al.}}

\titlerunning{Fundamental Cosmology in the E-ELT Era}

\abstract{
The recent observational evidence for the acceleration of the universe demonstrates that canonical theories of cosmology and particle physics are incomplete---if not incorrect---and that new physics is out there, waiting to be discovered. The most exciting task for the next generation of astrophysical facilities is therefore to search for, identify and ultimately characterise this new physics. Here we summarise ongoing work by CAUP's Dark Side Team aiming to identify optimal observational strategies for forthcoming facilities. The discussion is centred around the E-ELT (which will play a key role in this endeavour), but synergies with other ground and space-based facilities are also briefly considered. With the expected major improvements in the sensitivity of standard probes and entirely new ones such as the redshift drift (which the E-ELT, the SKA and possibly other facilities will measure) a new generation of precision consistency tests of the standard paradigm will become possible.
}

\maketitle{}

\section{Introduction}

In the middle of the XIX century Urbain Le Verrier and others mathematically discovered two new planets by insisting
that the observed orbits of Uranus and Mercury agreed with the predictions of Newtonian physics. The first of these---which we now call Neptune---was soon observed by Johann Galle and Heinrich d'Arrest. However, the second (dubbed Vulcan) was never found. We now know that the discrepancies in Mercury's orbit were a consequence of the fact that Newtonian physics can't adequately describe Mercury's orbit, and accounting for them was the first success of Einstein's General Relativity.

Over the past several decades, cosmologists have mathematically discovered two new components of the universe---which we have called dark matter and dark energy---but so far these have not been directly detected. Whether they will prove to be Neptunes or Vulcans remains to be seen but even their mathematical discovery highlights the fact that the standard $\Lambda$CDM paradigm, despite its phenomenological success, is at least incomplete.

Something similar applies to particle physics, where to some extent it is our confidence in the standard model that leads us to the expectation that there must be new physics beyond it. Neutrino masses, dark matter and the size of the baryon asymmetry of the universe all require new physics, and, significantly, all have obvious astrophysical and cosmological implications. Recent years have indeed made it clear that further progress in fundamental particle physics will increasingly depend on progress in cosmology.

After a quest of several decades, the recent LHC evidence for a Higgs-like particle \citep{atlas,cms} finally provides
strong evidence in favour of the notion that fundamental scalar fields are part of Nature's building blocks. A pressing
follow-up question is whether the associated field has a cosmological role, or indeed if there is some cosmological
counterpart.

It goes without saying that fundamental scalar fields already play a key role in most paradigms of modern cosmology.
Among others they are routinely invoked to describe period of exponential expansion of the early universe (inflation), cosmological phase transitions and their relics (cosmic defects), the dynamical dark energy which may be powering
the current acceleration phase, and the possible spacetime variation of nature's fundamental couplings.

Even more important than each of these paradigms is the fact that they don't occur alone: whenever a scalar field plays one of the above roles, it will also leave imprints in other contexts that one can look for. For example, in realistic models of inflation, the inflationary phase ends with a phase transition at which cosmic defects will form (and the energy scales of both will therefore be unavoidably related). More importantly, in the context of this workshop, in realistic models of dark energy, where the dark energy is due to a dynamical scalar field, this field will couple to the rest of the model and lead to potentially observable variations of nature's fundamental couplings; we will return to this point later in this contribution. Although this complementary point is often overlooked, it will be crucial for future consistency tests.

\section{Varying fundamental couplings}

Nature is characterised by a set of physical laws and fundamental dimensionless couplings, which historically we have assumed to be spacetime-invariant. For the former this is a cornerstone of the scientific method (it's hard to imagine how one could do science at all if it were not the case), but for the latter it is only a simplifying assumption without further justification. These couplings determine the properties of atoms, cells, planets and the universe as a whole, so it's remarkable how little we know about them. We have no 'theory of constants' that describes their role in physical theories or even which of them are really fundamental. If they vary, all the physics we know is incomplete.

Fundamental couplings are indeed expected to vary in many extensions of the current standard model. In particular, this will be the case in theories with additional spacetime dimensions, such as string theory. Interestingly, the first generation of string theorists had the hope that the theory would ultimately predict a unique set of laws and couplings for low-energy physics. However, following the discovery of the evidence for the acceleration of the universe this claim has been pragmatically replaced by an 'anything goes' approach, sometimes combined with anthropic arguments. Regardless of the merit of such approaches, experimental and observational tests of the stability of these couplings may be their best route towards a testable prediction.

It goes without saying that a detection of varying fundamental couplings will be revolutionary: it will immediately prove that the Einstein Equivalence Principle is violated (and therefore that gravity can't be purely geometry) and that there is a fifth force of nature. But even improved null results are important and useful. The simple way to understand this is to realise that the natural scale for cosmological evolution of one of these couplings (driven by a fundamental scalar
field) would be Hubble time. We would therefore expect a drift rate of the order of $10^{-10}$yr${}^{-1}$. However, current local bounds, coming from atomic clock comparison experiments, are 6 orders of magnitude stronger \citep{rs208}.

Recent astrophysical evidence suggests a parts-per-million spatial variation of the fine-structure constant $\alpha$ at low redshifts \citep{webb}; although no known model can explain such a result without considerable fine-tuning, it should also be said that there is also no identified systematic effect that can explain it. One possible cause for concern (with these and other results) is that almost all of the existing data has been taken with other purposes in mind, whereas this kind of measurements needs customised analysis pipelines and wavelength calibration procedures beyond those supplied by
standard pipelines. This is, of course, one of the reasons for the ongoing ESO UVES Large Programme, whose first results are discussed in the contributions of P. Molaro and H. Rahmani in these proceedings.

In the short term the PEPSI spectrograph at the LBT can also play a role here, and in the longer term a new generation of high-resolution, ultra-stable spectrographs like ESPRESSO (for the VLT) and ELT-HIRES, which have these tests as a key science driver, will significantly improve the precision of these measurements and should be able to resolve the current controversy. A key technical improvement will be that ultimately one must do the wavelength calibration with laser frequency combs.

In theories where a dynamical scalar field yields varying $\alpha$, the other gauge and Yukawa couplings are also expected to vary. In particular, in Grand Unified Theories the variation of $\alpha$ is related to that of energy scale of Quantum Chromodynamics, whence the nucleon masses necessarily vary when measured in an energy scale that is independent of QCD (such as the electron mass). It follows that we should expect a varying proton-to-electron mass ratio, $\mu=m_p/m_e$. Obviously, the specific relation between $\alpha(z)$ and $\mu(z)$ will be highly model-dependent, but this very fact makes this a unique discriminating tool between competing models.

It follows from this that it's highly desirable to identify systems where various constants can be simultaneously
measured, or systems where a constant can be measured in several independent ways. Systems where combinations of constants can be measured are also interesting, and may lead to consistency tests \citep{FJMM1,FJMM2}. These points are illustrated in the contributions of M. Juli\~ao and A.M. Monteiro in these proceedings.

In passing, let us also briefly comment on other probes of varying constants. The CMB is in principle a very clean one, but in most simple models a parts per million variation of at redshifts a few leads to variations at redshift $z\sim1089$ that are below the sensitivity of Planck. However, these studies do have a feature of interest, namely that they lead to  constraints on the coupling between the putative scalar field and electromagnetism, independently (and on a completely different scale) from what is done in local tests, as illustrated in \citet{erminia}; another example is provided in M. Martinelli's contribution to these proceedings. Compact objects such as solar-type stars and neutron stars have also been leading to interesting constraints \citep{jpv,angeles}.

\section{Dynamical dark energy and varying couplings}

Observations suggest that the universe is dominated by an energy component whose gravitational behaviour is quite similar to that of a cosmological constant. Its value is so small that a dynamical scalar field is arguably a more likely explanation. Such a field must be slow-rolling (which is mandatory for $p<0$) and be dominating the dynamics around the present day. It follows that if the field couples to the rest of the model (which it will naturally do, unless some symmetry is postulated to suppress the couplings) it will lead to potentially observable long-range forces and time dependencies of the constants of nature.

In models where the degree of freedom responsible for the varying constants also provides the dark energy, the redshift of the couplings is parametrically determined, and any available measurements (be they detections of null results) can be used to set constraints on combinations of the scalar field coupling and the dark energy equation of state. See the contributions of R. Thompson and P. Vielzeuf in these proceedings for illustrations of this point. One can show that ELT-HIRES will either find variations or rule out---at more than 10 sigma---the simplest classes of these models (containing a single linearly coupled dynamical scalar field).

However, this is not all. Standard observables such as supernovae are of limited use as dark energy probes, both because they probe relatively low redshifts and because to ultimately obtain the required cosmological parameters one effectively needs to take second derivatives of noisy data. A clear detection of varying $w(z)$ is crucial, given that we know that $w\sim-1$ today. Since the field is slow-rolling when dynamically important (close to the present day), a convincing detection of a varying $w(z)$ will be tough at low redshift, and we must probe the deep matter era regime, where the dynamics of the hypothetical scalar field is fastest.

Varying fundamental couplings are ideal for probing scalar field dynamics beyond the domination regime \citep{amendola}: such measurements can presently be made up to redshift $z\sim4$, and future facilities such as the E-ELT may be able to significantly extend this redshift range. Importantly, even null measurements of varying couplings can lead to interesting constraints on dark energy scenarios. ALMA, ESPRESSO and ELT-HIRES can realise the prospect of a detailed characterisation of dark energy properties all the way until $z\sim4$, and possibly beyond. In the case of ELT-HIRES, a reconstruction using quasar absorption lines is expected to be more accurate than using supernova data (its key advantage being huge redshift lever arm), See P. Pedrosa's contribution to these proceedings, as well as \citet{amendola}, for further details.

Dark energy reconstruction using varying fundamental constants does in principle require a mild assumption on the field coupling, but there are in-built consistency checks, so that inconsistent assumptions can be identified and corrected. Explicit examples of incorrect assumptions that lead to observational inconsistencies can be found in \citet{moi1} and P. Vielzeuf's contribution to these proceedings.

It's important to keep in mind that the E-ELT will also contribute to the above task by further means. First and foremost there is the detection of the redshift drift signal. This is a key driver for ELT-HIRES, and possibly---at a fundamental level---ultimately the most important E-ELT deliverable. Indeed, as shown in \citet{moi1}, having the ability to measure the stability of fundamental couplings and the redshift drift with a single instrument is a crucial strategic advantage. (Nevertheless, it should also be said that other facilities such as PEPSI at the LBT, the SKA and ALMA may also be able do measure the redshift drift.) Additionally, the ELT-IFU (in combination with JWST) should find Type Ia supernovas up to a redshift $z\sim5$. An assessment of the impact of these future datasets on fundamental cosmology is currently in progress. Interesting synergies are also  expected to exist between these ground-based spectroscopic methods and Euclid, which need to be further explored.

\section{Consistency tests}

Whichever way one finds direct evidence for new physics, it will only be trusted once it is seen through multiple independent probes. This was manifest in the case of the discovery of the recent acceleration of the universe, where the supernova results were only accepted by the wider community once they were confirmed through CMB, large-scale structure and other data. It is clear that history will repeat itself in the case of varying fundamental couplings and/or dynamical dark energy. It is therefore crucial to develop consistency tests, in other words, astrophysical observables whose behaviour will also be non-standard as a consequence of either or both of the above.

The temperature-redshift relation,
\begin{equation}
T(z) = T_0(1+z)
\end{equation}
\noindent is a robust prediction of standard cosmology; it assumes adiabatic expansion and photon number conservation, but it is violated in many scenarios, including string theory inspired ones. At a phenomenological level one can parametrise deviations to this law by adding an extra parameter, say
\begin{equation}
T(z) = T_0(1+z)^{1-\beta}
\end{equation}

Our recent work \citep{Avgoustidis} has shown that forthcoming data from Planck, ESPRESSO and ELT-HIRES will lead to much
stronger constraints: Planck on its own can be as constraining as the existing (percent-level) bounds, ESPRESSO can improve on the current constraint by a factor of about three, and ELT-HIRES will improve on the current bound by one order or magnitude. We emphasise that estimates of all these gains rely on quite conservative on the number of sources (SZ clusters and absorption systems, respectively) where these measurements can be made. If the number of such sources increases, future constraints can be correspondingly stronger.

The distance duality relation,
\begin{equation}
d_L = (1+z)^2d_A
\end{equation}
\noindent is an equally robust prediction of standard cosmology; it assumes a metric theory of gravity and photon number conservation, but is violated if there's photon dimming, absorption or conversion. At a similarly phenomenological
level one can parametrise deviations to this law by adding an extra parameter, say
\begin{equation}
d_L = (1+z)^{2+\epsilon}d_A
\end{equation}
\noindent with current constraints also being at the percent level, and improvements are similarly expected from Euclid,
the E-ELT and JWST.

In fact, in many models where photon number is not conserved the temperature-redshift relation and the distance duality relation are not independent. With the above parametrisations it's easy to show \citep{Avgoustidis} that
\begin{equation}
\beta=-\frac{2}{3}\epsilon
\end{equation}
\noindent but one can in fact further show that a direct relation exists for any such model, provided the dependence is in redshift only (models where there are frequency- dependent effects are more complex). This link allowed us to use distance duality measurements to improve current constraints on $\beta$, leading to
\begin{equation}
\beta= 0.004\pm0.016
\end{equation}
\noindent which is a $40\%$ improvement on the previous constraint. With the next generation of space and ground-based experiments, these constraints can be further improved (as discussed above) by more than one order of magnitude.

In models where the degree of freedom responsible for the varying constants does not provide (all of) the dark energy, the link to dark energy discussed in the previous section no longer holds. However, has shown in \citet{moi1}, such wrong assumptions can be identified through (in)consistency tests. For example, it has been shown in \citet{moi4} that in Bekenstein-type models one has
\begin{equation}
\frac{T(z)}{T_0}=(1+z)\left[\frac{\alpha(z)}{\alpha_0}\right]^{1/4}\sim(1+z)\left(1+\frac{1}{4}\frac{\Delta\alpha}{\alpha}\right)
\end{equation}
\begin{equation}
d_L(z)\sim d_A(z)(1+z)^2\left(1+\frac{3}{3}\frac{\Delta\alpha}{\alpha}\right)
\end{equation}

Interestingly these also hold for disformal couplings (but not for chameleon-type models, where the powers of $\alpha$ are inverted), These effects are relevant for the analysis of Planck data: a parts-per-million $\alpha$ dipole leads, in this class of models, to a micro-Kelvin level dipole on the CMB temperature, in addition to the usual milli-Kelvin one due to our motion.

Note that even if this degree of freedom does not dominate at low redshifts it can still bias cosmological parameter estimations, For example. in varying-$\alpha$ models the peak luminosity of Type Ia supernovas will depend on redshift. This scenario is currently being studied in more detail---see M. Martinelli's contribution in these proceedings for some preliminary results.

Now, if photon number non-conservation changes $T(z)$, the distance duality relation, etc, this may lead to additional biases, for example for Euclid. In \citet{moi4} we have quantified how these models weaken Euclid constraints on cosmological parameters, specifically those characterising the dark energy equation of state. Our results show that Euclid can, even on its own, constrain dark energy while allowing for photon number non-conservation. Naturally, stronger constraints can be obtained in combination with other probes. Interestingly, the ideal way to break a degeneracy involving the scalar-photon coupling is to use $T(z)$ measurements, which can be obtained with ALMA, ESPRESSO and ELT-HIRES (which, incidentally, may nicely complement each other in terms of redshift coverage). It may already be possible to obtain some useful constraints from Planck clusters, and these will be significantly improved with a future PRISM mission.

Last but not least, the role of redshift drift measurements as a consistency test cannot be over-emphasised \citep{moi1}. Standard dark energy probes are geometric and/or probe localised density perturbations, while the redshift drift provides a unique measurement of the global dynamics \citep{sandage,loeb,liske}. It does not map out our (present-day) past light-cone, but directly measures evolution by comparing past light cones at different times. Therefore it provides an ideal probe of dark sector in deep matter era, complementing supernovas and constants. In fact, as recently shown in \citet{moi3}, its importance as a probe of cosmology does not stem purely from its intrinsic sensitivity, but also from the fact that it is sensitive to cosmological parameters that are otherwise hard to probe (in other words, it can break some key degeneracies). One illustrative example \citep{moi3} is that the CMB is only sensitive to the combination $\Omega_m h^2$, while the redshift drift is sensitive to each of them.

\section{Conclusions}

We have highlighted the key role that will be played by forthcoming high-resolution ultra-stable spectrographs in fundamental cosmology, by enabling a new generation of precision consistency tests. The most exciting and revolutionary among these is clearly the redshift drift, which is a key driver for ELT-HIRES, but may also be within the reach of other facilities, like PEPSI (at the LBT), SKA or even ALMA (although no sufficiently detailed studies exist for these at present).

Finally, let us point out that the ELT will enable further relevant tests, including tests of strong gravity around the galactic black hole (through ELT-CAM), and astrophysical tests of the Equivalence Principle, which were not discussed in this contribution. Interesting synergies with other facilities, particularly ALMA and Euclid, remain to be fully explored.
 
\begin{acknowledgements}
We acknowledge the financial support of grant PTDC/FIS/111725/2009 from FCT (Portugal). CJM is also supported by an FCT Research Professorship, contract reference IF/00064/2012. We are also grateful to the staff at the Sexten Center for Astrophysics (Gabriella and Gabriella) for the warm hospitality during the meeting.
\end{acknowledgements}

\bibliographystyle{aa}

\end{document}